\documentclass[useAMS,usenatbib]{mn2e}
\usepackage{epsfig,amsmath,amssymb,amsfonts,mathrsfs,latexsym,graphicx}
\begin{document}

\author[Maurer \& Watts]{Immanuel Maurer$^1$ and Anna L. Watts$^2$
\\Max Planck Institut f\"ur
  Astrophysik, Karl-Schwarzschild-Str.1, 85741 Garching, Germany\\
$^1$ maurer@mpa-garching.mpg.de, $^2$anna@mpa-garching.mpg.de}

\title[Ignition latitude and X-ray burst shape]{Ignition latitude and the shape of Type I X-ray bursts}

\maketitle

\begin{abstract}
The shape of the lightcurve during the rising phase of Type I X-ray
bursts is determined by many factors including the ignition latitude,
the accretion rate, and the rotation rate of the star. We develop
a phenomenological model of the burst rise process and 
show that simple measures of the burst morphology can be robust
diagnostics of ignition latitude and burning regime.  We apply our
results to the large sample of bursts from the Low Mass X-ray Binary
4U 1636-536, and find evidence for off-equatorial ignition for many of
the bursts.  We argue that such behaviour may be associated with the
transition from hydrogen to helium ignition at accretion rates a few percent of
Eddington.  We show that this model can also explain variations in the
detectability of burst oscillations, and discuss the implications for
other burst sources.    
\end{abstract}


\maketitle

\section{Introduction}
\label{intro}

Neutron stars in Low Mass X-ray Binaries (LMXBs) accrete matter from
their low mass companions via Roche lobe 
overflow.  Nearly half of these systems show Type I X-ray bursts,
thermonuclear flashes caused by rapid unstable burning of the
accumulating hydrogen or helium after it settles and is compressed on
the neutron star 
surface. A
typical X-ray burst light curve has a rapid rise (less than 10 s), followed by a longer
decaying tail persisting for seconds to minutes as the star's surface
cools.  

The basic properties of Type I X-ray bursts can be understood in terms of
the stability of hydrogen and helium burning at different accretion
rates \citep{FHM, FL, Bild, B, NH, W, CNc, PBT}.  The boundaries between the
different burning regimes depend on the local accretion rate
$\dot{m}$ (accretion rate per unit area). Below a critical accretion
rate $\dot{m}_\mathrm{c1} \sim 
1$\% of the Eddington rate $\dot{m}_\mathrm{Edd}$, bursts are
triggered by unstable hydrogen burning.  Above
this level, hydrogen burns stably via the hot CNO cycle, and bursts are
triggered by unstable helium burning once a critical column depth is
reached.  For $\dot{m} > \dot{m}_\mathrm{c1}$ but below a second
critical rate $\dot{m}_\mathrm{c2}$, expected to be a few percent of $\dot{m}_\mathrm{Edd}$,
the hydrogen at ignition depth should burn before helium ignition,
leading to a pure helium flash.   At 
higher accretion rates, $\dot{m} > \dot{m}_\mathrm{c2}$, hydrogen does
not burn completely
before the burst is triggered, leading to various classes of mixed
H/He bursts. Bursting activity will cease when the accretion rate is
sufficiently high, $\dot{m} > \dot{m}_\mathrm{c3}$, that helium
burning is also stable. The precise 
levels of the critical accretion rates $\dot{m}_\mathrm{c1},
\dot{m}_\mathrm{c2}$, $\dot{m}_\mathrm{c3}$, are not known precisely and will vary depending
on factors such as the metallicity of the accreted material and the
heat flux from the deep crust and core \citep{AJ, FL, BBR, NH}.  

There is now observational evidence for bursts in all three
classes.  The global accretion rate $\dot{M}$ can be inferred from the
X-ray luminosity, and this should (under the assumption that accretion
is spherically
symmetric) predict $\dot{m}$ and hence the burning regime.  Furthermore, within
a given burning regime, burst rate should rise with accretion
rate.  For a large number of sources, however, these
expectations are not borne out.  Transition accretion rates can differ
from the predicted values by up to an order of magnitude, and for many
sources burst rate actually falls as $\dot{M}$ increases \citep{C,
  GMCPH}.  Why this should be the case is not yet clear but it may
involve either additional stable burning processes \citep{B95} or
non-spherical accretion and slow fuel spread (so that local accretion
rate $\dot{m}$ differs from that inferred from $\dot{M}$)
\citep{B}. Alternatively it may well be that X-ray luminosity is not a
good predictor of accretion rate \citep{HV}.  For a more
in-depth discussion of these issues and other burst properties we
refer the reader to the recent articles by \citet{SB} and \citet{GMCPH}.

In this paper we concentrate on one particular aspect of burst physics - the
properties of the burst rise.  The shape and time scale of the rising
portion of the X-ray light curve is controlled by several factors: the
point at which the burning layer ignites, the emission profile of any
burning point after ignition (set by the nuclear heating and cooling
time scales); and the speed at which the burning front propagates
across the stellar surface.  Ignition is likely to occur at a point
and spread, rather than occurring across the whole star
simultaneously, because of the discrepancy between the very short
burning time scales and the much longer accretion time scale
\citep{SH}. Asymmetric initialization is also thought
necessary to explain the detection during the rise of burst oscillations, variations
in brightness on the surface of the neutron star that are modulated by
the rapid stellar spin \citep{SB}. 

On a realistic neutron star, ignition location is unlikely to be
random due to the various factors that break spherical symmetry.
Accretion flow may 
not be spherically symmetric - it may occur through an
equatorial boundary layer \citep{IS} or by channeling onto the
magnetic poles (in which case the star may manifest as an X-ray
pulsar). Whether either of these issues can affect local ignition
conditions or lead to preferred ignition locations is not clear, since
estimates suggest that fuel should spread rapidly across the stellar
surface between bursts.  Another factor that will certainly have an
impact is stellar rotation.  Most of the neutron stars that show
Type I X-ray bursts are thought to be rapidly rotating, which reduces
the effective gravity at the
equator. Centrifugal effects, coupled with the deformation of the
neutron star due to the rapid rotation, combine to reduce the
effective gravity at the equator as compared to the
poles. This results in a local accretion rate that is higher at the
equator \citep{SLU}.  The 
column depth required for ignition 
is achieved more rapidly, and ignition should occur
preferentially at this latitude.  In fact there are
accretion rates where off-equator ignition is still expected.
\citet{CNa} considered the situation at high 
accretion rates $\dot{m} \approx \dot{m}_\mathrm{c3}$ where helium
burning is on the verge of stability.  There will be a range of
accretion rates (the more rapid the rotation, the larger this range)
where $\dot{m} > \dot{m}_\mathrm{c3}$ at the equator but not at other
latitudes.  Although not discussed by \citet{CNa}, a similar region of
off-equator ignition is to be expected at $\dot{m} \approx
\dot{m}_\mathrm{c1}$ (the transition to stable hydrogen burning).

Once ignition has occurred, the nuclear burning processes determine
the emission from a given point.  The burning layer will expand during
a convective phase, and there is a delay before radiative processes
take over and the light curve starts to rise.  Light curves from a
single point (ignoring spreading effects) have been generated
by a number of authors \citep{T, AJ, FTWL, TWWL, W, WBS}.  The shape,
time scales and strength of the single point 
light curve can vary substantially depending on factors such
as the burning regime and the composition of the accreted material.
There are as yet no simple analytic models for this process that take into
account all of the relevant parameters.  

At the accretion rates of relevance to most burst sources, the burning
front is expected to propagate by deflagration \citep{FW, HF, B95}
rather than by detonation \citep{FWa, Z}.  In the simplest picture,
spreading speed is set by the rate at which heat is transported
across the burning front (by convective processes).  \citet{SLU} have
since shown that rapid rotation will also play a significant
role:  interaction between the uplift (vertical expansion) of burning
material and a strong Coriolis force can act to slow spreading.  The
degree of asymmetry in the spread of the burning front is also
 relevant to the detectability of burst oscillations in the
rise, if they are caused by a growing hot spot.  

In this paper we attempt the first systematic examination of how these various
factors interact to affect the shape and time scale of the rising
portion of the light curve.  Previous studies in this area have focused
on small samples of bursts such as the rare 
multi peaked bursts \citep{BSa, BSb}, or bursts from the accreting
millisecond pulsars \citep{BSc}.  In this study we
adopt a much broader remit, motivated by the wide variety of shapes
exhibited by the
bursts of the LMXB 4U 1636-536.  This particular source is an excellent
candidate for this type of study: there are over 120 bursts from this
source in the {\it Rossi X-ray Timing Explorer} (RXTE) archive, and it
lies in a binary with relatively well-constrained properties.   By
comparing the
burst properties with the results of parameterized simulations, we show that
simple measures of the burst shape can be profound diagnostics of ignition latitude and
burning regime.  

The paper is structured as follows.  In Section \ref{obs} we
define simple measures of burst morphology and classify the bursts
of 4U 1636-536 accordingly.  Section \ref{sims} gives details of the
parameterized simulations that we carried out to generate model
light curves, and looks at the effects on shape using the same simple
measures.  In Section \ref{disc} we compare the results of our
simulations with the data, and consider the implications in terms of
ignition point and burning regime.  We present a model that can
explain our results (subject to the assumptions inherent in our
simulations), and extend our analysis to other sources to test its
feasibility.  We conclude in Section \ref{conc}.

\section{Observational data analysis}
\label{obs} 

The bursting LMXB 4U 1636-536 is a persistent atoll source at a
distance of $\approx 6$ kpc 
\citep{GPMC} in a 3.8
hour binary orbit with a low mass blue star, V801 Ara.  The
binary inclination $i$ is in the range
$36^\circ-60^\circ$ \citep{CA}.  The bursts have been studied extensively
with both EXOSAT \citep{LVT} and RXTE \citep{GMCPH}.  Burst
oscillations at 581 Hz are seen in both normal bursts and
superbursts, the latter being longer bursts triggered by unstable
carbon burning \citep{SZSWL, SM}.   We include in our study all of the
bursts covered by the RXTE burst catalogue 
\citep{GMCPH}.  The catalogue includes over 1000 bursts observed by the RXTE
Proportional Counter Array (PCA)
from December 30th 1995 to October 29th 2005, 123 of which originate
from 4U 1636-536.  

\begin{figure} 
\begin{center}
\includegraphics[width=8cm, clip]{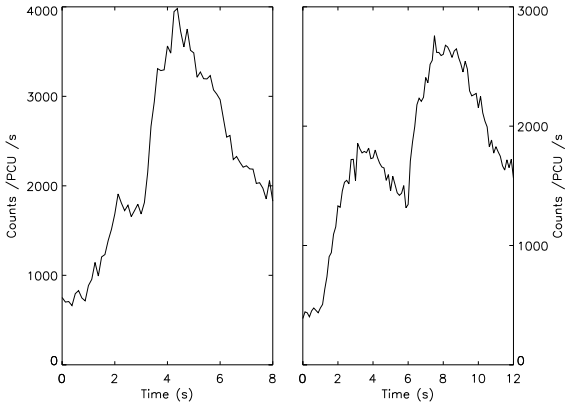}
\end{center}
\caption{Examples of kinked and multi-peaked structure in the burst
  rise for bursts from 4U 1636-536.  The left panel shows Burst 18 for
  this source (using the numbering system from \citet{GMCPH}):  this
  burst shows a kinked rise.  The right panel shows Burst 40, which
  has a strongly multi-peaked rise.}
\label{kinked}
\end{figure}

In order to study the shape and time scales of the bursts' rises we started by
constructing light curves from the PCA data, using Standard 1 data
(0.125 s resolution) and using
all available energy channels.  For most of the bursts this results in
a smooth and monotonically increasing light curve.  For a small group
of bursts with peak fluxes below $5 \times 10^{-9}$ ergs cm$^{-2}$
s$^{-1}$ (bursts 11, 67, 71, 76, 78, 87, 88, 91, 118 and 122 for
  this source, using the numbering system of \citet{GMCPH}), the
  light curve is not smooth, so we exclude these bursts from our
  analysis.  Burst 117 is also excluded because the rise is truncated.  We also exclude for the moment bursts that are either kinked or
multi-peaked when plotted on a coarser (0.5s) time resolution:
that is to say bursts
where the gradient of the lightcurve drops to zero or below 
during the rise.  This group includes the strongly multi-peaked
bursts (40, 48, 56 and 111) identified as such by \citet{GMCPH} as
well as a group that are either kinked 
or weakly multi-peaked (bursts 5, 18, 90, 110 and 113), see
Figure \ref{kinked}.  We will discuss these bursts in more detail in
Section \ref{disc}.   

\begin{figure}  
\begin{center}
\includegraphics[width=9cm, clip]{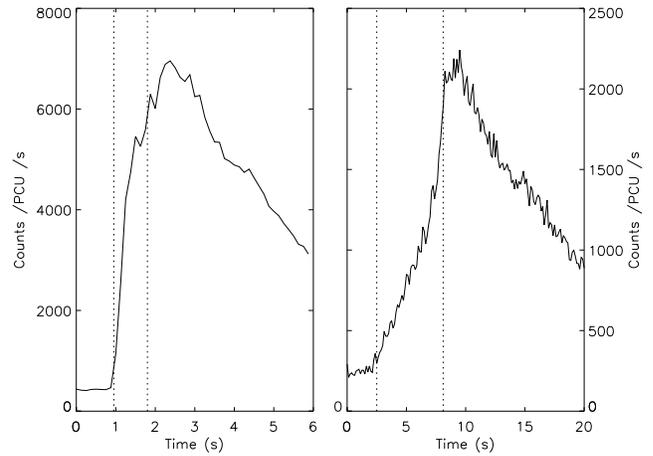}
\end{center}
\caption{Light curves at 0.125 s resolution for two
  bursts from 4U 1636-536, one convex and one concave.  The left panel
  shows Burst 
  1 and the right panel Burst 52 (numbers from \citet{GMCPH}). The
  dashed lines mark the
  portion of the burst rise where the count rate lies between 10 and 90
percent of the maximum (corrected for the pre-burst emission).  It is
this portion of the curve that we use to calculate convexity.}   
\label{f3}
\end{figure}

\begin{figure} 
\begin{center}
\includegraphics[width=9cm, clip]{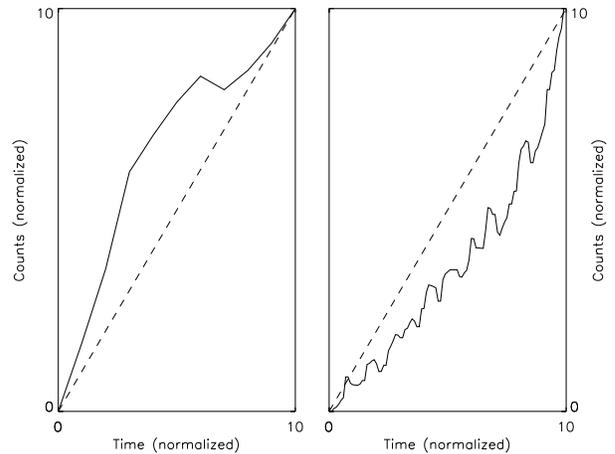}
\end{center}
\caption{Re-normalized light curves for the two bursts from Figure
  \ref{f3}.  As before, the left panel shows Burst 1, and the right
  panel Burst 52.  Convexity $\mathcal{C}$, defined in equation
  (\ref{conv}), is the integrated area above or below the diagonal dashed
  lines, with areas above the line (convex bursts) taken to be
  positive and areas 
  below the line (concave bursts) negative.  For Burst 1, we find
  $\mathcal{C} = 14.8 $, and for Burst 52  $\mathcal{C} = -20.5$.}
\label{normalized}
\end{figure}

In analyzing the light curves of the remaining 103 bursts
 we considered several different ways of
quantifying the shape.  The measure that we found to be most powerful,
which we call convexity ($\mathcal{C}$),
measures the degree to which the light curve is concave or convex.
The physical significance of this parameter will become clear
in Section \ref{sims}, where we find that it can be used to diagnose
the latitude at which the burst ignites.  Convexity $\mathcal{C}$ is calculated
as follows.  We define the burst rise to be the interval where the
count rate rises from 10 to 90 percent of the maximum count rate
(corrected for the pre-burst persistent emission).  We define rise
time $\tau_\mathrm{R}$ as the duration of that interval. In order
to compare the shape of bursts of different durations and peak
count rates we then
normalize both quantities so that they rise
from 0 to 10 in dimensionless units.  This process is illustrated in Figures
\ref{f3} and \ref{normalized}.  Taking $c_i$ as the re-normalized count rate in each bin, and $x_i$ as
the identity function 
(shown as a diagonal line in Figure \ref{normalized}), we define
convexity $\mathcal{C}$ as 

\begin{equation}
\mathcal{C} = \sum_{i = 0}^N (c_i - x_i)  \Delta t 
\label{conv}
\end{equation}
where N is the number of re-normalized time bins and $\Delta t$ is the
re-normalized time bin size.  Convexity is effectively the integrated area of the
curve above or below the diagonal line - areas above the line being
positive and areas below negative.  For highly convex bursts, where most points lie above
the diagonal, $\mathcal{C}$ is positive, and for highly concave bursts
it is negative.  Our choice of re-normalization means that
$\mathcal{C}$ lies in the range -50 to 50.  

\begin{figure} 
\begin{center}
\includegraphics[width=9cm, clip]{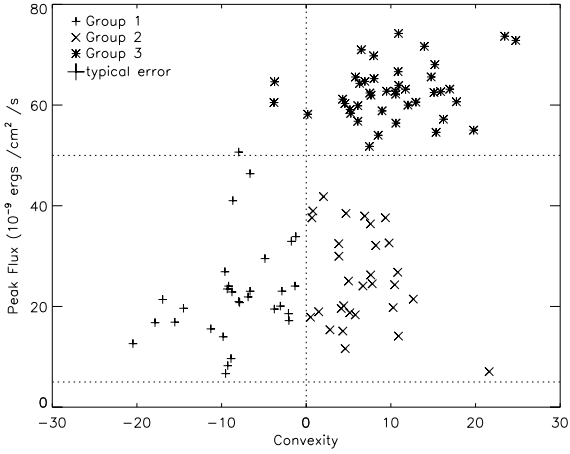}
\end{center}
\caption{Peak flux for the bursts of 4U 1636-536 (from \citet{GMCPH}) against
  convexity.  The bursts divide naturally into two
  groups based on peak flux (above or below $50\times 10^{-9}$
  ergs cm$^{-2}$ s$^{-1}$).  For convenience in later discussions we
  also split the low peak flux bursts into two groups depending on
  whether convexity is positive or negative.  Only
  three bursts (those in the upper left quadrant) sit outside this
  simple grouping.  For these we use the presence or absence of radius
  expansion to determine group membership:  Group 3 bursts show radius
  expansion while those in Group 1 do not.  The lower dotted line
  shows the minimum peak flux for which we calculate convexity
  ($5\times 10^{-9}$ ergs cm$^{-2}$ s${-1}$).}  
\label{f1}
\end{figure}

Figure \ref{f1} plots the peak flux $F_\mathrm{p}$ of the bursts against
convexity.  At high peak fluxes almost all of the bursts have
$\mathcal{C} > 0$, whereas at low peak fluxes we find equal numbers
of bursts with both 
positive and negative convexities.  For convenience in later
discussions, we assign the bursts to three broad groups, as indicated
in the Figure.  Bursts with $F_\mathrm{p} < 50\times 10^{-9}$
  ergs cm$^{-2}$ s$^{-1}$ are assigned to Group 1 ($\mathcal{C} < 0$)
  or Group 2 ($\mathcal{C}>0$), while the bright photospheric radius
  expansion (PRE)
  bursts with $F_\mathrm{p} >  50\times 10^{-9}$
  ergs cm$^{-2}$ s$^{-1}$ form Group 3.   There are no PRE bursts in Group
1, and only one in Group 2\footnote{Burst 16, identified by \citet{GPMC} as a
PRE burst with an 
anomalously low peak flux, due most likely to the presence of
hydrogen.}. 

\begin{figure} 
\includegraphics[width=9cm, clip]{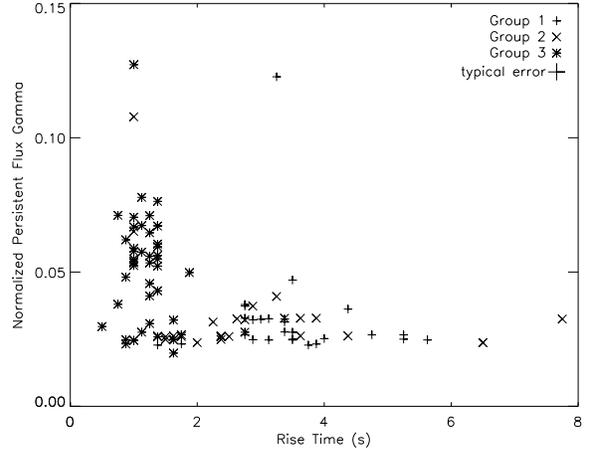}
\caption{Normalized persistent flux $\gamma$ (as a fraction of the
  Eddington rate, taken from \citet{GMCPH}) against rise time
  $\tau_\mathrm{R}$ for the 4U 1636-536 bursts.  Groups 1 and 2 cluster in
  the range $\gamma \approx 0.02 - 0.04$. Group 3 bursts are
  seen at all values of $\gamma$ and 
  dominate at higher $\gamma$.  Groups 1 and 2 
  have longer rise 
  times than Group 3 (for which $\tau_\mathrm{R} \approx 1$ s).  There
  are only four Group 1/2 bursts at $\gamma > 0.05$.  Two (partly obscured) have
  $\gamma \approx 0.07$.   The Group 2
  burst at $\gamma = 0.108$ is Burst 16, one of the anomalous PRE bursts
  identified in \citet{GPMC} (the other, Burst 18, is kinked and is
  not included on this plot). The Group 1 burst at the highest
  $\gamma$ is Burst 19:  the burst rise shape is unusual, being almost
kinked.}  
\label{f2}
\end{figure}

\begin{figure} 
\begin{center}
\includegraphics[width=9cm, clip]{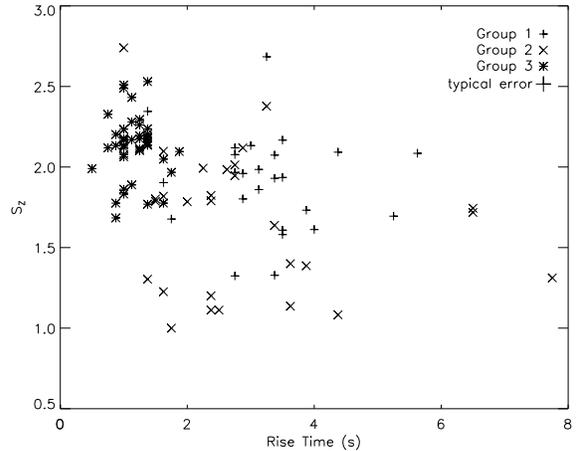}
\end{center}
\caption{Colour-colour diagram position $S_z$ \citep{GMCPH} against
  rise time.  Groups 1 and 2 dominate at lower $S_z$, while Group 3
  dominates at higher values.}  
\label{szrisetime}
\end{figure}

Figures \ref{f2} and \ref{szrisetime} show the relationship between
accretion rate and burst rise time\footnote{Note that we use a different
  definition of rise time to \citet{GMCPH}.  These authors define rise
  time as the time taken for the count rate to rise from 25\% to 90\%
  of the peak rate.} $\tau_\mathrm{R}$ for the bursts. \citet{GMCPH} give
two parameters that can be used to    
estimate the accretion rate at the time of the burst.  The first, the
normalized persistent flux
$\gamma$, is the ratio of the persistent X-ray flux to the Eddington flux. This is thought to be a reasonable
estimate of the global accretion rate $\dot{M}$ as a fraction of the
Eddington rate $\dot{M}_\mathrm{Edd}$. The second measure, $S_z$, measures position in a
colour-colour diagram: $S_z$ is thought to increase as accretion rate
rises.   Both measures show the same trend.  The bursts in Groups 1
and 2 cluster at low accretion rates (a few percent of
Eddington) and have longer rise times than those in Group 3.  The
Group 3 bursts have short rise times and dominate at higher accretion rates.  

Table \ref{bprops} summarizes the mean properties of the three
groups of bursts.  In addition to the parameters previously defined,
we include burst fluence $E_b$ (the integrated flux during the whole
burst) and the burst time scale $\tau_b$ (the ratio of burst fluence to
peak flux), using data from \citet{GMCPH}.  Group 3 bursts have higher
fluences and shorter 
time scales, whereas properties for Groups 1 and 2 are similar. We also indicate the percentage of bursts in each group for which
\citet{GMCPH} state that oscillations are detected
during the burst rise phase.  Oscillations are far more prevalent in
Group 3 than in Groups 1 and 2, as previously noted for this source by
\citet{MCGS}.  

\begin{table*}
\begin{tabular}{|l|c|c|c|}
\hline
Property  & Group 1 & Group 2 & Group 3\\
\hline
Number of bursts in group & 31 & 30 & 42 \\
$\tau_R$ (s) & 3.3 $\pm$ 1.0  & 2.8 $\pm$ 1.6& 1.2 $\pm$ 0.3\\
$\tau_b$ (s) & 10.8 $\pm$ 4.9 & 13.7 $\pm$ 6.6& 7.7 $\pm$ 2.2\\
$E_b$ (10$^{-6}$ ergs cm$^{-2}$) & 0.24 $\pm$ 0.02 & 0.33 $\pm$ 0.03 &
0.48 $\pm$ 0.02\\
$\gamma$  & 0.033 $\pm$ 0.003 & 0.033 $\pm$ 0.003 & 0.052 $\pm$ 0.003\\
$S_z$ &  1.91    $\pm$ 0.06  &  1.69    $\pm$ 0.08 & 2.13  $\pm$ 0.03 \\
$F_\mathrm{p}$ (10$^{-9}$ ergs cm$^{-2}$ s$^{-1}$) & 22.7 $\pm$ 1.8& 25.5
$\pm$ 1.9 & 62.7 $\pm$ 0.8 \\
 \% bursts with oscillations in rise & 7\% & 23\% & 36\% \\
\hline
\end{tabular}
\caption{Mean properties for the three groups of bursts.  Most
  parameters are taken from \citet{GMCPH} apart from rise time, where
  we use the definition given in this paper.}
\label{bprops}
\end{table*}

\section{Simulations}
\label{sims}

To understand what might cause the variations in shape and time scale
seen in the observational data, we developed a simple phenomenological
model of the burst rise process and ran parameterized simulations to
generate light curves.  Our model consists of three main
elements:  a time-dependent temperature profile used to describe the
emission from each point on the surface after ignition; a velocity
model that describes the propagation of 
the burning front across the neutron star surface; and a light curve
generation routine that models the propagation of photons from the neutron star
surface towards a distant observer.   

To generate light curves we use the Oblate Schwarzschild (OS)
approximation of \citet{MLCB} to model 
relativistic light-bending, Doppler shifts and 
gravitational redshift.  The OS model, which takes into account
rotation-induced oblateness, is more appropriate for very rapidly
rotating neutron stars than the more usual Schwarzschild + Doppler
approximation \citep{PGie}.  Assuming that burst oscillation frequency
is a good measure of stellar spin\footnote{For the two accreting
  millisecond pulsars that also show burst 
oscillations, the burst oscillation frequency is at or very close to
the known spin frequency \citep{CM, SMSI}.  The situation for the non-pulsing
LMXBs is less clear (there are some differences in burst oscillation
properties compared to the pulsars, see \citet{WS}). However, in all of
the suggested models (Section \ref{intro}) burst oscillation
frequency lies within a few Hz of the spin frequency.}, 4U 1636-536
rotates at $\approx 580$ Hz.  The associated rotational deformation is
a few percent, depending on the assumed mass and nuclear equation of
state, leading to a small but noticeable influence on the light curve
\citep{MLCB}.  We neglect both special relativistic time delay and the
additional time delay experienced by initially inwardly propagating
photons, since these delays are much smaller ($\sim 10^{-4}$ s, \citet{PBel}) than the time bins we
consider ($\sim 0.1$ s).  We specify stellar mass $M$ and equatorial radius $R_\mathrm{eq}$ and
then compute the deformed spherical surface using the OS model. 

To
start the burst we specify an initial small burning area and
then track the propagation of the burning front across the star.  The
stellar surface is divided into a grid of patches with area $\sim$ 
0.1 km$^2$, and we consider a patch to be ignited as soon as the burning
front reaches the centre point of the patch.  Once a patch has started
burning, we need to specify how its emission varies
with time.  As discussed in Section
\ref{intro}, there are various numerically-generated single point
emission models, but no simple analytic models.  In this study we follow
\citet{BSa, BSb} and assume that the
temperature of the burning front follows the following profile after ignition:

\begin{eqnarray}
T & = & T_0 + (T_1 - T_0)[1 -
\exp(-t/t_\mathrm{lr})],  ~~~~  t \le t_m \nonumber \\
& = &  T_m \exp(- t/t_\mathrm{ld}), ~~~~ t \ge t_m
\label{temp}
\end{eqnarray}
where $t_m$ is the time at which the temperature reaches its
maximum $T_m = T_0 + 0.99(T_1-T_0)$. The time scale $t_\mathrm{lr}$
sets the time scale at which the temperature increases, while
$t_\mathrm{ld}$ sets the time scale on which it decays.    
Unburnt patches are assumed to have a temperature $T_0$ until
ignition.  The parameters in this model depend primarily on the
composition of the burning material, which will vary with on
accretion rate (see Section \ref{intro}).  Bursts which are
helium-rich, for example, would be expected to have shorter timescales than
those which contain a higher fraction of hydrogen.  Our parameter
space must therefore be wide enough to take into account the expected
level of variation.  One follow-on question is whether this
exponential temperature model remains valid across 
all  
burning regimes.  It has been used successfully in detailed spectral
modelling of
bursts at different accretion rates by \citet{BSa,BSb}, but ultimately
one would like to see this confirmed by detailed nuclear physics
calculations.  

\begin{figure} 
\begin{center}
\includegraphics[width=9cm, clip]{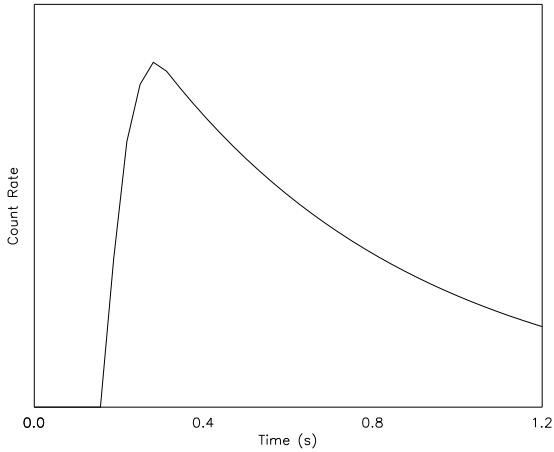}
\end{center}
\caption{A typical lightcurve for a single patch on the NS
  surface before taking into account propagation effects. Note that
  the y-axis is linear. The rise
  time scale $t_\mathrm{lr} = 0.1$ s, and the
  decay time $t_\mathrm{ld} = 6$ s.  All single point lightcurves for
  the assumed temperature profile are intrinsically convex:
  $\mathcal{C} \sim 10$ for $t_\mathrm{lr} \sim 1$ s, falling as
  $t_\mathrm{lr}$ gets shorter.}  
\label{singlepoint}
\end{figure}

We then assume black body emission at the specified temperature from
each patch.  In Figure \ref{singlepoint} we show the typical single patch
light curve.  The rise portion is similar in shape to the bursts shown in
\citet{W}.  The decay portion does differ from that seen in some
of the cases studied by \citet{W}, but because we are focusing on the
rise we never reach the points late in the decay phase where the
difference would become relevant.  This relatively simple analytic
formulation is therefore a reasonable approximation to a more
realistic calculation, for the short rising phase that we are
studying. To this we can then apply various different beaming
functions.   

In adopting this emission model we have made two assumptions.
We have assumed that each point on the star will have the
same emission profile.  This may not be the case, since emission could
vary (as a result, say, of inhomogeneities in fuel deposition
or composition).   We have also neglected any delay between ignition
and emission.  If a convective zone develops there will be a delay
as the zone grows before radiative processes become a more efficient
form of transport, and radiation starts to escape from the photosphere
\citep{WBS}.  If this delay is position-independent
then there will be no effect on the overall shape of the lightcurve;
there could however be some position-dependence here that would
contribute to variations in emission profile across the star.
 
In modeling the spreading of the burning front, we use the expression
for burning front speed, $v_\mathrm{flame}$,
developed by \citet{SLU}.  This
speed depends on various nuclear burning parameters as well as the
(poorly known)
strength of frictional coupling between the top and bottom of the
burning ocean.  In
general the velocity is latitude dependent (faster at the equator), because of the role
of the Coriolis
force.  \citet{SLU} show that the burning front speed varies as

\begin{equation}
v_\mathrm{flame} \sim \Big[\frac{gh_\mathrm{hot}}{t_n}
\frac{1/t_\mathrm{fr} + \eta/t_n}{f^2 + 
    (1/t_\mathrm{fr} + \eta/t_n)^2} 
    \Big]^{1/2}.
\label{spreadspeed}
\end{equation}
In this expression, $g$ is the acceleration due to gravity at the surface, $h_\mathrm{hot}$
is the scale height of the hot burnt material, and $f = 2\Omega
\cos\theta$ is the Coriolis parameter, $\Omega$ being the angular frequency
the star and $\theta$ the latitude.  The parameter $t_\mathrm{fr}$ is
the time scale for frictional coupling between the top and bottom
layers of the burning ocean. The parameter $t_n$ is the nuclear time
 scale of the thermonuclear burning (set by the composition of the
 burning material), and $\eta$ is a constant of
 order unity.   In the case of weak frictional coupling ($t_\mathrm{fr} \gg
 t_n, 1/f$)\footnote{Note that this approximation breaks
   down in a very 
small region near the equator, where $1/f \rightarrow \infty$.
In this case we have to use the full expression for velocity.} 

\begin{equation}
v_\mathrm{flame} = \frac{v_p}{\cos \theta}
\label{vlowf}
\end{equation}
where the velocity at the pole, $v_p \sim \sqrt{gh_\mathrm{hot}}/2\Omega
t_n$. If frictional effects are stronger ($t_\mathrm{fr} \le t_n$), behaviour depends on
how $t_\mathrm{fr}$ compares to $1/f$.  Maximum flame speed is reached
when  $t_\mathrm{fr} = 1/f$. In this case

\begin{equation}
v_\mathrm{flame} = \frac{v_\mathrm{pm}}{\sqrt{\cos \theta}}
\label{vmax}
\end{equation}
with $v_\mathrm{pm} \sim \sqrt{gh_\mathrm{hot}/4 \Omega t_n}$.
In the high frictional coupling limit ($t_\mathrm{fr} \ll 1/f$) the latitude
dependence disappears, and

\begin{equation} 
v_\mathrm{flame} \sim \sqrt{\frac{gh_\mathrm{hot}t_\mathrm{fr}}{t_n}}.
\label{vhighf}
\end{equation}    

The shape of the light curve depends on a whole host of parameters:
however some have a larger effect than others.  We start by
presenting results for a baseline scenario, where some of the
parameters are fixed.  Later in
this section we will show that varying 
these additional parameters has only a limited effect on our findings.  In the
baseline scenario we assume a neutron star mass $M = 1.4 M_\odot$ and
an equatorial radius $R_\mathrm{eq} = 12$ km.  We set the spin rate of the neutron star to 580 Hz,
and assume a binary inclination $i = 50^\circ$, in the middle of the
range inferred for 4U 1636-536 by \citet{CA}.  In the spreading
speed model we assume that frictional coupling is weak (equation
\ref{vlowf}), so that flame speed is inversely proportional to
$\cos\theta$.  We start our simulations by setting an initial burning
area with radius 1 km, similar to the expected width of the flame
front \citep{SLU}.  In the temperature model (equation \ref{temp}) we
fix the following parameters: $T_0 = 1$ keV, $T_1 = 2.8$
keV, and $t_\mathrm{ld} = 6$ s (as used by \citet{BSa}).  We also
assume beaming $\propto 
\cos\psi$, where $\psi$ is the angle from the normal to the surface.  

The parameters that we vary are the ignition latitude
$\alpha_\mathrm{ign}$, the polar velocity $v_p$ and the temperature
rise time scale $t_\mathrm{lr}$.  We vary ignition latitude in
$30^\circ$ steps from $0^\circ$ (north pole) to $180^\circ$ (south
pole).   The parameters $v_p$ and $t_\mathrm{lr}$ both depend on the
composition of the burning material, which will vary with accretion
rate as the fraction of hydrogen changes.  We choose ranges for these
 parameters wide enough to encompass the level of variation that
might be expected as we move from pure helium bursts to hydrogen-rich
bursts \citep{W, WBS}.  We consider $v_p = (1, 2, 4,
8, 16) \times 10^5 $ cm/s (for velocities outside this range
our simulations give rise times are either much shorter or much longer than those
observed). In the temperature profile we consider 
$t_\mathrm{lr}
= 0.001, 0.1, 0.5, 1, 2$ s, a range wide enough to cover the values generated in
the single point light curve models of, for example, \citet{WBS}.  

\begin{figure*}
\includegraphics[width=18cm, clip]{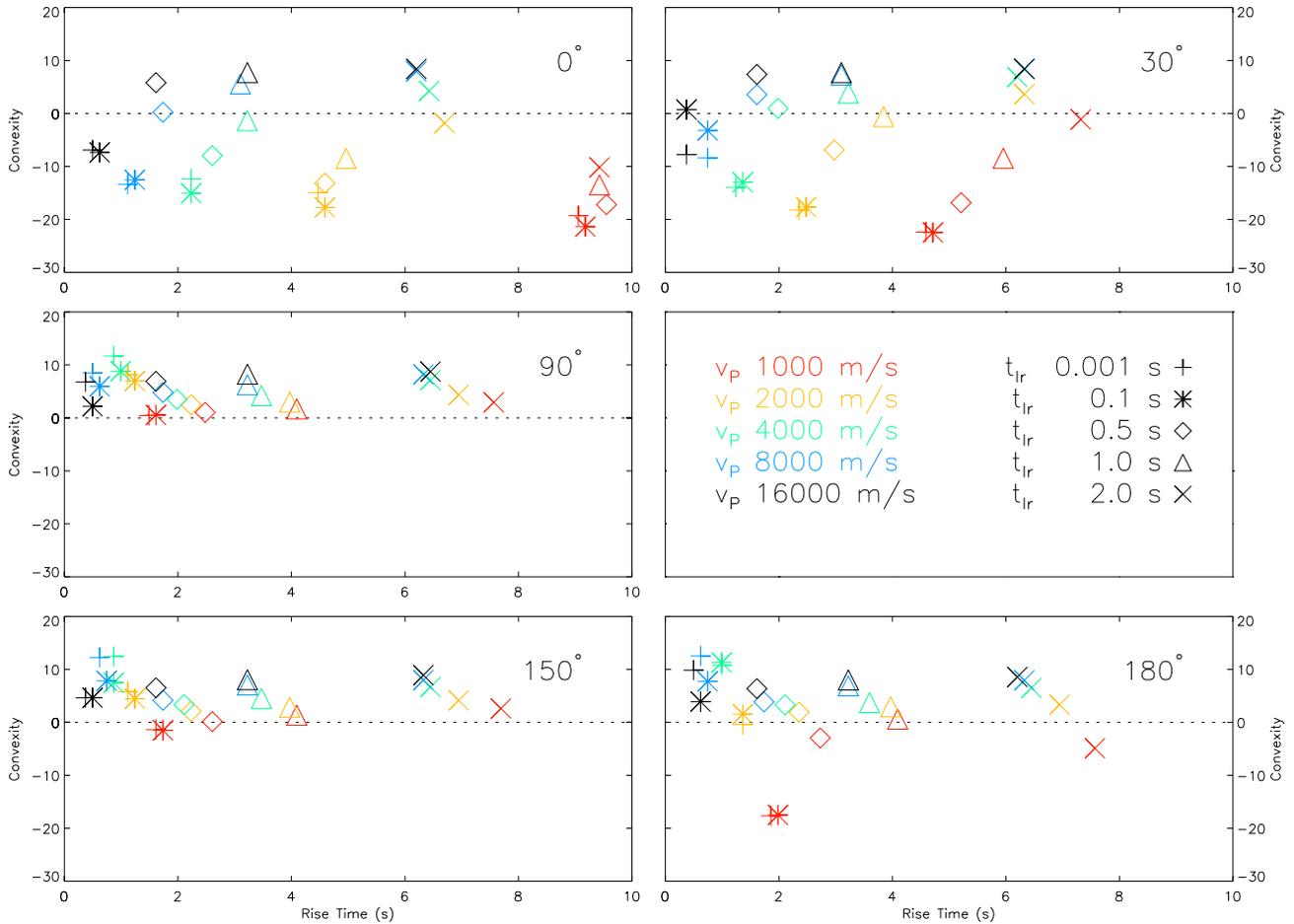}
\caption{Convexity $\mathcal{C}$ against rise time
  $\tau_R$ for simulated bursts, showing the effect of changing $v_p$ and
  $t_\mathrm{lr}$.  The range of $v_p$ and $t_\mathrm{lr}$ studied
  covers the variation expected as the composition of the burning
  material changes from pure helium to a mix containing a substantial
  fraction of hydrogen. Each panel shows a different ignition latitude:
  $\alpha_\mathrm{ign} = 0^\circ$ (north pole), $\alpha_\mathrm{ign} =
  30^\circ$, $\alpha_\mathrm{ign} = 90^\circ$ (equator),
  $\alpha_\mathrm{ign} = 150^\circ$, and $\alpha_\mathrm{ign} =
  180^\circ$ (south pole).  The cases $\alpha_\mathrm{ign} =
  60^\circ$ and $120^\circ$ are not shown but are very similar to the
  equatorial case.  For ignition at the equator, convexity is
  always positive.  Convexity decreases as
  ignition moves towards the poles, and can become negative.  Negative
convexity bursts are much more common for ignition in the northern
hemisphere than the southern. Rise times are shortest near the equator, increasing
as ignition point moves towards the poles, with north pole ignition bursts
showing longer rise times than south pole ignition bursts. The
difference in behavior between the poles reflects the fact that we
have set $i = 50^\circ$.}
\label{latmany}
\end{figure*}

Figure \ref{latmany} shows the results from our simulations. It is
clear that ignition
latitude has a dramatic effect on both convexity and rise times.  We
find the following key result: bursts ignited on the equator always have $\mathcal{C} > 0$, for the
full range of $v_p$ and $t_\mathrm{lr}$ considered.  Convexity
decreases as ignition latitude moves towards the poles, and 
can eventually become negative at high enough latitudes; negative convexity being much more
easily achieved when ignition starts in the northern hemisphere than
in the southern.  Rise times do vary slightly with ignition latitude
but depend more strongly on spreading speed and the temperature rise
timescale.   The differences between
northern and southern hemisphere ignition are due to the fact that we
are observing the star from a northern hemisphere vantage point ($i =
50^\circ$).  For bursts ignited at the south pole we do not observe
the initial slow, concave, part of the rise that is visible to us if the burst
ignites at the north pole.  A south pole ignited burst therefore appears to have a higher convexity, purely by virtue of our viewing angle.  

Let us now consider whether our results are robust when we vary the
parameters that were held constant in our previous simulations: 
neutron star mass and radius, inclination angle, and the size of the initial
burning area.  We also vary the remaining temperature parameters, which are
functions of composition and hence accretion rate. To do
this we took a representative sample of bursts (Table \ref{btests}), varied
the other parameters, and compared the results to the baseline
scenario with the same ignition latitude, $v_p$ and $t_\mathrm{lr}$.

\begin{table}
\begin{tabular}{l|c|c|c|}
\hline
Burst ID & $\alpha_\mathrm{ign}$ ($^\circ$) & $v_p$ ($\times 10^5$
cm/s) & $t_\mathrm{lr}$ (s)\\
\hline
1 & 0  & 2 & 0.1 \\ 
2 & 0  & 2 & 0.5 \\
3 & 0  & 4 & 0.1 \\
4 & 0  & 4 & 0.5  \\
5 & 90 & 2 & 0.1 \\
6 & 90 & 2 & 0.5  \\
7 & 90 & 8 & 0.1  \\
8 & 90 & 8 & 0.5  \\
9 & 180 & 2 & 0.1  \\ 
10 & 180 & 2 & 0.5   \\
11 & 180 & 4 & 0.1   \\
12 & 180 & 4 & 0.5   \\
\hline
\end{tabular}
\caption{Fixed parameters for the twelve bursts that we used in
  testing the effects of the parameters that were held constant in the
baseline scenario. For the equatorial ignition bursts we considered a
wider range of $v_p$ because these values were a better fit for the
observed rise times. }
\label{btests}
\end{table}

We considered neutron star masses in the  
range $M = 1$ to $1.8 M_\odot$ and equatorial radii $R_\mathrm{eq}$ 10
to 14 km, in line with reasonable estimates for a range of equations
of state \citep{LP}.  We varied inclination from $i =
0^\circ$ to $90^\circ$ (a range much larger than that inferred
for
4U 1636-536, but which will allow us to extend our results to other
sources in Section \ref{disc}).  The radius of the initial burning
region was varied from 200 m to 1 km. We took $t_\mathrm{ld}$ in the
range 4-15 s, 
and tested the effect of increasing both $T_0$ and $T_1$ by a factor of
5.  We also included the effect of the RXTE spacecraft
response on our simulated bolometric light curves. To do this we
generated typical 
PCA response functions using some of the 
observed bursts analyzed in Section \ref{obs}.  We then used XSPEC to
fold our simulated light curves through the response functions, and
re-computed convexity and rise times for the folded light curves. 

\begin{figure} 
\includegraphics[width=9cm,  clip]{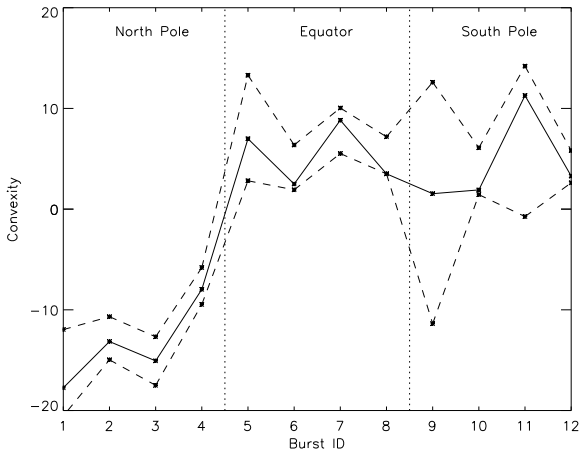}
\caption{Convexity for the 12 test bursts (Table \ref{btests}). The black line shows
  the baseline scenario. The dashed lines show the maximum
  deviation from this scenario when we vary $t_\mathrm{ld}$, $M$, $R$,
  ignition patch size and include the RXTE response function, for inclinations in the range $40^\circ
  - 60^\circ$ (the range appropriate for 4U 1636-536).}  
\label{devallcon}
\end{figure}

\begin{figure} 
\includegraphics[width=9cm, clip]{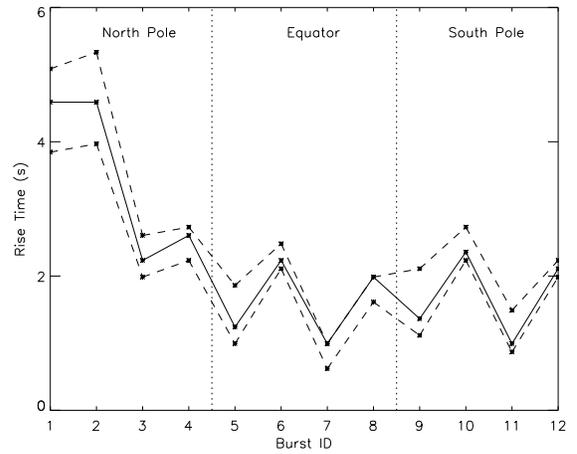}
\caption{As for Figure \ref{devallcon}, but showing the effect on rise time.}
\label{devallrt}
\end{figure}

\begin{figure} 
\includegraphics[width=9cm, clip]{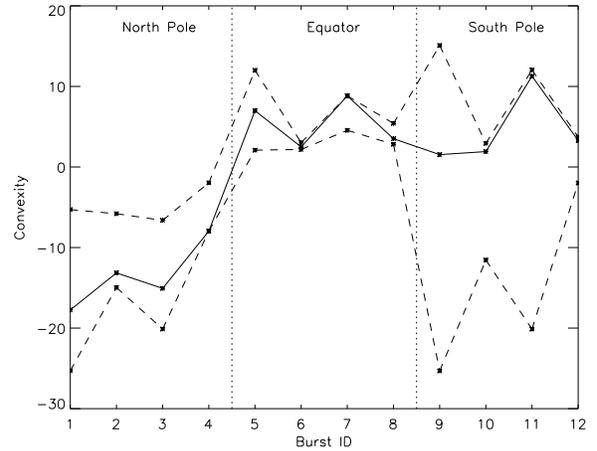}
\caption{As for Figure \ref{devallcon}, for inclinations in the range
  $0^\circ - 90^\circ$.  At $i = 90^\circ$, as
  expected, north and 
  south polar ignition bursts become identical.  As inclination falls,
  convexity increases:  at $i = 0^\circ$, for example, the values are
  too high to 
  explain the observations from 4U 1636-536. }
\label{devallcon2}
\end{figure}

Figures \ref{devallcon} and \ref{devallrt} show the deviations in (respectively)
convexity, and burst rise time, as compared to the standard scenario,
for inclinations in the range $40^\circ - 60^\circ$ (the range
appropriate for 4U 1636-536).  The results of our previous analysis
are clearly robust.  In Figure \ref{devallcon2} we show results for a
much wider range of inclinations. The range of variation is much
wider, but again our main conclusions stand: negative convexity requires
off-equatorial ignition.  

We also checked the effects of several other parameters that are not
shown in the Figures.  One thing that we tested was the effect of
changing the beaming factor.  We first tried removing the beaming
factor: this had very little effect on bursts ignited in the northern
hemisphere or at the equator.  Burst ignited in the southern
hemisphere, however, were more prone to having negative convexities.
We also tried the more physically motivated grey atmosphere model
\citep{ML}.  In this model, the
temperature of each patch 

\begin{equation}
T^4 (\psi) = \frac{3}{4} T^4_\mathrm{eff} \left[\frac{2}{3}\cos \psi + 0.7\right]
\end{equation}
where $T_\mathrm{eff}$ is the effective
temperature of the emitting layer and $\psi$ is the angle from the
normal.  The effect on convexity was extremely small compared to the
baseline scenario (as might be expected, since both have the same
$\psi$ dependence). 

We also varied the latitude-dependence of spreading speed.  In
the baseline scenario we assumed that we were in the low frictional
coupling regime, where latitude-dependence is strongest
($v_\mathrm{flame} \propto 1/\cos\theta$).   By the intermediate friction regime,
$v_\mathrm{flame} \propto 1/\cos^{1/2}\theta$, while in the highest
friction regime the dependence on latitude vanishes. We therefore
considered two additional models, $v_\mathrm{flame} = v_p/ \cos^{1/2}
\theta$ and $v_\mathrm{flame} = v_p$.  Interestingly both of these
models failed to reproduce the observed convexities.  The constant
velocity case produced no negative convexity bursts at all.  The
intermediate friction case did generate some bursts with $\mathcal{C}
< 0$ but the values were not sufficiently negative to match the
observations.

\section{Discussion}
\label{disc}

\subsection{Ignition latitude and burning regime}

We start by summarizing the most important results from the preceding
sections.  In our analysis of the bursts of 4U 1636-536, we identified
three populations: one group with higher peak fluxes and two groups
with lower peak fluxes.  The high peak flux group dominate at higher
accretion rates and have short rise times.  All but two of the members
of this group have positive convexity.  The lower peak flux bursts divide
into two groups of similar size: one with positive convexity and one
with negative convexity.  Rise times in this group are longer than for
the high peak flux group, suggesting a different burning regime. 

In our simulations, both composition of the burning material (hydrogen
fraction) and ignition latitude have a strong influence on convexity
(Figure \ref{latmany}).  However only bursts that ignite near the
poles have negative 
convexities:  bursts ignited on or near the
equator always have positive convexity.  In addition, bursts ignited
near the north pole always have lower convexity than those ignited
near the south pole.   We can
therefore draw the following conclusion\footnote{Subject, of course,
  to the assumptions in our modeling.  Negative convexity bursts
  could, for example,
be generated by equatorial ignition if the single point lightcurves
were concave rather than convex (see Figure
\ref{singlepoint}). Although we are not aware of any 
physical motivation for such an assumption, we ran several simulations
with concave single point lightcurves to study the possibility.  In these
simulations, however, we could no longer generate positive convexity
bursts with long 
rise times. To explain the presence of both positive and negative
convexity bursts at low accretion rates we would therefore require
both types of single point lightcurve (temperature profile) to operate
simultaneously.}. The Group 1 bursts, which
have negative convexity, must be triggered by polar
ignition.  In the rest of the Section
we will outline a simple model that might explain why this should be
the case.

As discussed in Section \ref{intro}, there may be a region
of polar ignition at high accretion rates at the point
where He ignition is transitioning to stability \citep{CNa}.  However, the same
arguments should also apply at much lower accretion rates ($\dot{m} 
\approx \dot{m}_\mathrm{c1}$), where H ignition is on the verge of stability. We
therefore propose the following model.  At the lowest accretion rates,
the bursts of 4U 1636-536 are mixed H/He bursts, triggered by unstable
H ignition.  The star is however close to the transition where H burning
stabilizes, in the range of accretion rates where off-equatorial
ignition is preferred.   We expect
this range to be reasonably broad for 4U 1636-536 because of the
rapid stellar rotation.  As accretion rate increases H ignition 
stabilizes\footnote{The transition is actually rather more
  complex. Between the regime where H ignition triggers mixed H/He
  bursts, and the higher accretion rate regime where H burning is
  stable, there exists a range of accretion rates where hydrogen burns
  unstably via weak flashes 
  \citep{PBT, CNb}.  The flashes do not
  trigger associated He burning, so are faint and most likely
  undetectable above the persistent accretion luminosity.  The weak
  hydrogen flashes may also show a move from equatorial to polar
  ignition as accretion rate rises.}, and there 
is a transition to He-ignited bursts. Above the
transition ignition moves back to the equator.  

Within this picture the Group 1 and Group 2 bursts are mixed H/He
bursts triggered by 
off-equatorial H ignition.  We would expect similar numbers of bursts
to be triggered in the northern and southern hemispheres.  We
therefore suggest that most of the Group 1 bursts, which have negative
convexities, ignite in the northern hemisphere,
while most of the Group 2 bursts ignite in the
southern hemisphere (for which positive convexities are more
likely)\footnote{Some of the Group 2 bursts could ignite on the
  equator, as this would also give rise to positive convexity.  This
  might be expected at the lowest accretion rates, and could explain
  the apparent excess of Group 2 bursts at the lowest values of $S_z$,
  see Figure \ref{szrisetime}.}.
The two groups should have similar peak 
fluxes and rise times, in accordance with our observations.  The
properties of the 
bursts in these groups are in accordance with those expected for mixed
H/He bursts triggered by H ignition: low peak fluxes, rise
times of a few seconds, and durations longer than 10s. The value of
$\gamma$ at which the transition takes place ($\gamma \approx 0.03$)
is perhaps a little 
higher than expected (in \citet{PBT}, for example, H ignition no
longer triggers mixed H/He bursts above $\approx$ 1\% Eddington), but
the precise values of the transitions will depend on factors such as
heating from the deep crust, and $\gamma$ is not a perfect measure of
accretion rate.  

The Group 3 bursts would in this picture be triggered by He ignition on
the equator.  At accretion rates immediately above the transition one
would expect the bursts to be nearly pure He, with the amount of H
involved in the bursts increasing as accretion rate rises.  The
properties of the Group 3 bursts are in line with those expected for
He-dominated bursts:  high peak fluxes (with radius expansion), 
rise times $\lesssim 1$ s, and durations $\lesssim 10$ s.  We note that a
transition to short rise time radius expansion bursts at a few percent
of the Eddington rate was also seen in the EXOSAT data \citep{L}.

A transition from mixed
H/He burning triggered by H ignition, to He burning, should result in
a drop in burst rate and an increase in alpha (the ratio of the energy
released by stable burning between the bursts to the energy released
in bursts). If our interpretation is correct this should occur when
the accretion rate is  $\approx 3-5$ \% $\dot{M}_\mathrm{Edd}$, and
this is indeed what is observed (see Figure 16 of
\citet{GMCPH}).  This picture might also
explain some unusual features of the PRE 
bursts for this source.  \citet{GPMC} found that while most of the PRE
bursts reached the Eddington limit for pure He, two had lower peak fluxes,
requiring some H in the mix.  These two
exceptional PRE bursts occur, 
as discussed in Section \ref{obs}, at the highest accretion rates.  At
these rates pure He bursts are no longer likely and bursts should have some
mixed H/He character again, reducing the peak flux reached by the
PRE bursts. As shown by \citet{GC}, there can be a substantial
percentage of H in the  mix before the Eddington limit starts to fall
below that expected for pure He.  

The values of convexity that we measured for the bursts of 4U 1636-536
ranged from -20 for the Group 1 bursts up to +20 for the Group 3
bursts.  Although our simulations generated bursts 
with negative convexities in the right range, convexities $> 10$ were
harder to generate (Figure \ref{latmany}).  One reason for this is
that our simulations did not take into account PRE (which all of the
Group 3 bursts show).  PRE tends to flatten the top of the lightcurve:
when we include this in our simulations it results in an increase in
convexity of sufficient magnitude to explain the discrepancy.
Changing the parameters from the baseline scenario can also increase
convexity (compare Figures \ref{latmany} and \ref{devallcon}).  A rise in
peak temperature, for example (as might be expected for He-rich
bursts) increases convexity.  

By comparing the observed values of convexity and rise time with those
generated by the simulations, we can infer the range of
$t_\mathrm{lr}$ and $v_p$ required to explain the observations
(comparing Figures \ref{f1}-\ref{f2} and Figure \ref{latmany}).  For
the Group 3 bursts, for example, we need $t_\mathrm{lr} \lesssim 0.5$ s
independent of $v_p$,
while for the Group 2 (and hence also the Group 1) bursts
we require 
$t_\mathrm{lr} \gtrsim 0.1$ s.  We can 
therefore ask whether the values that we infer are in line with the
values predicted for the suggested burning regimes. The single point lightcurve models of \citet{WBS} predict a
temperature rise 
timescale of $\sim 0.01$ s for He bursts, rising to $\sim 0.1$ s as H
fraction increases. The timescales derived by \citet{W} using
multizone models are slightly longer.  The limits that we derive are
therefore broadly compatible with these models.  What is harder to
check is the validity of the inferred values of $v_p$.  Spreading
speed (equation \ref{spreadspeed}) depends on the nuclear timescale,
the scale height of 
the burnt atmosphere and the strength of frictional coupling in the
burning layers.  Naively one would expect the nuclear timescale to be
related to the temperature rise timescale (although the degree to
which convection develops could skew this relationship).  The
dependence of the frictional coupling on the burning regime, however,
is very poorly understood.  Without a better understanding of the
burning and spreading process, it is difficult to say whether the values
of $v_p$ suggested by our simulations accord with the values expected for the
different burning regimes.    

\subsection{Accretion disk effects}

One issue that we have not considered in our modelling is the role of
the accretion disk.  An optically thick accretion disk extending down
to the stellar surface could obscure the southern hemisphere of the
star.  At the lowest accretion rates, the disk is likely to be
truncated sufficiently far from the star for our unobscured models to
be valid \citep{DG}.  As
accretion rate increases, however, the inner disk 
 is expected to move in towards the stellar surface, obscuring
the southern hemisphere.

Radiation pressure from a bright burst may be able to push the disk
back, revealing the southern 
hemisphere once more \citep{STH}. Even if the southern hemisphere were
obscured, however, there would be little impact on our results.
Bursts ignited in the northern hemisphere would still have negative
convexities, and bursts ignited at the equator would still be
positive, although peak countrates would be lower and apparent rise
times shorter than in our simulations.  The largest impact would be
on bursts ignited in the 
southern hemisphere:  these would only be visible to us once the
burning front traversed the equator, so would all have positive
convexities.  

\subsection{Burst oscillations}
\label{oscns}

In Table \ref{bprops} we summarized the detectability of burst
oscillations, as reported by \citet{GMCPH}, for the rising phase of
bursts from 4U 1636-536.  The detectability criteria used were based
on power exceeding a certain threshold in short time bins (0.25s).
Oscillations were most likely to be detected in Group 3 and then in
Group 2.  Group 1 bursts were far less likely to show oscillations.  Can
this be explained by the pattern of flame spreading if the bursts in
Groups 1 and 2 
result from polar ignition, while those in 
Group 3 result from equatorial ignition?  Detectability will depend on
relative amplitude of the asymmetry, the length of time for which it
persists, and the overall countrate.  A long-lived lower amplitude asymmetry may be more
detectable than a short-lived high amplitude asymmetry if the time
window used for the power spectral analysis is long.    

\begin{figure} 
\includegraphics[width=9cm, clip]{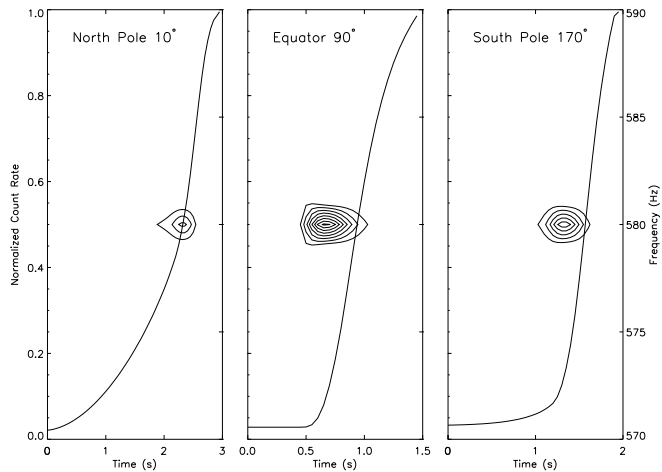}
\caption{Dynamical power spectra from simulated
  bursts with ignition starting just off the north pole
  ($\alpha_\mathrm{ign} = 10^\circ$), at the equator
  ($\alpha_\mathrm{ign} = 90^\circ$) and just off the south pole
  ($\alpha_\mathrm{ign} = 170^\circ$). Note that we start just off the
  poles because bursts ignited exactly at the pole spread in a totally
  axisymmetric fashion and there would be no burst oscillations.  For
  these simulations we used the baseline scenario with $v_p$ = 4000 m/s and
  $t_\mathrm{lr}$ = 0.1 s.  The peak countrate for the polar bursts is
  2000 cts/s while the
  peak countrate for the equatorial burst is 4000 cts/s (see the
  explanation in the text).  We
  use 0.25s time windows (to match those used for oscillation
  detection by \citet{GMCPH}), overlapping by 0.05s.  The contours
  start at a power of 5 and rise in increments of 5.  Note the
  variation in strength and duration.}
\label{latampev}
\end{figure}

We carried out a number of simulations to test whether our model is
compatible with the observations.  In Figure
\ref{latampev} we show dynamical power spectra from 
simulated bursts ignited at different latitudes.   To mimic the
differences between the different groups we assumed a higher peak
countrate for the equatorial ignition burst.  We find that
oscillations are most detectable in the bright equatorial ignition
bursts (our model for the Group 3 bursts). For the bursts with lower
peak countrates, 
oscillations are 
more detectable for bursts that ignite in the southern hemisphere.
This fits with our interpretation in which Group 2 bursts are ignited
near the south pole and Group 1 bursts near the north pole. 

\subsection{Multi-peaked and kinked bursts}

4U 1636-536, like many other neutron stars, shows
both multi-peaked bursts and bursts with kinks (at least two points of
inflection) in the
rise. This group includes those that we excluded from our analysis in
Section \ref{obs} as well as bursts with weaker kinks such as Burst
19 (see Figure \ref{f2}) and the two bursts with negative convexities
in Group 3 (Figure \ref{f1}).  The simulations described in Section
\ref{sims}, which involved 
single point ignition and subsequent smooth spread, produced no
simulated light curves with more than one point of inflection. 

We therefore extended
our simulations to investigate the effect of near
simultaneous ignition at equivalent latitudes in the northern and 
southern hemispheres.  We wanted to see whether a second burning front
igniting before it could be engulfed by a burning front spreading from
the other hemisphere could give a multi-peaked or kinked light curve.
We ran a series of simulations where north and south pole were ignited
either 
simultaneously or at intervals from 0.5 to 4s.  In no case were
we able to replicate either multi-peaked or kinked shapes.  

We conclude that some additional physics is still required to explain
the multi-peaked and kinked bursts.  \citet{BSa, BSb} put forward a
scenario in which multi-peaked bursts were caused by polar ignition in
one hemisphere, some kind of stalling on the equator, followed by the
burning front re-activating and propagating on towards the other
pole.  In \citet{WM} we argued against this model on the grounds that
polar ignition was only expected at the highest accretion rates \citep{CNa},
whereas the multi-peak bursts are actually seen at low and
intermediate rates.  However, as we have pointed out in this paper,
polar ignition is also likely to play a role at the transition to
stable H burning.  A mechanism involving polar ignition may therefore
still be responsible, provided that a viable stalling mechanism can be
identified.  Interestingly three of the multi-peaked bursts from 4U
1636-536 sit at the accretion 
rates where we are postulating polar ignition.  The fourth, which
looks rather different, however, sits at slightly higher rates \citep{WM}.

\subsection{Application to other sources}
\label{others}

\begin{figure}
\includegraphics[width=9cm,clip]{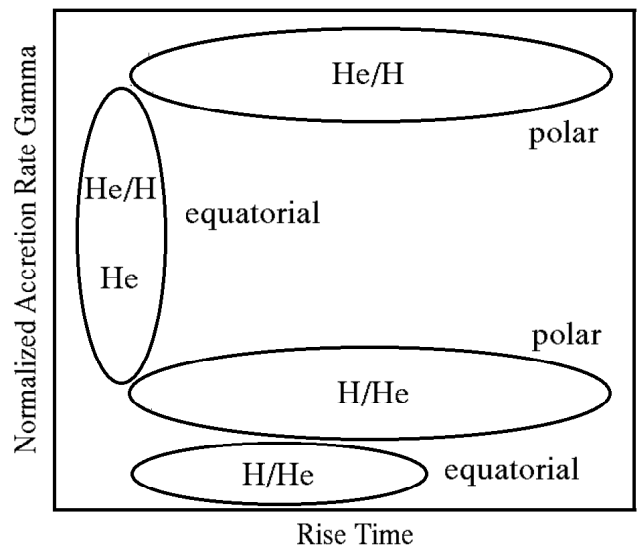}
\caption{A schematic of the type of behaviour we expect in a plot of
  $\gamma$ (a proxy for accretion rate) against burst rise time for a
  rapidly rotating neutron star:  in the text we refer to this as the F diagram.  At
  the lowest accretion rates, bursts are triggered by H ignition and
  ignite at the equator.  As accretion rate rises ignition moves
  off-equator and rise times increase, forming the lower horizontal
  stroke of the F.  When H burning stabilises, ignition (now by He)
  moves back to the equator and we move up the vertical stroke of the
  F.  Rise times should increase as the proportion of H in the bursts
  increases, and we may move onto the upper horizontal stroke of the F
  before He ignition moves off-equator at the highest accretion rates.} 
\label{fdiag}
\end{figure}

If ignition does indeed move towards the pole as we reach the upper
limits of the accretion rate where H ignition is still feasible - and
then back to the equator when He ignition begins - then we would
expect to see similar trends in other sources.  Assuming a favorable
inclination we would also expect to see positive convexity bursts at
the very lowest accretion rates (when H ignition should occur at the
equator), and negative convexity bursts at the very highest accretion
rates (when He ignition moves off-equator towards the pole).  Figure
\ref{fdiag} illustrates the type of behavior that we expect, and the
regions where polar ignition should occur.  In what follows we will
refer to this plot as the F diagram. 

We selected a number of additional sources that are thought to
accrete a mix of H/He from the RXTE burst catalogue
\citep{GMCPH}. We chose the sources that span
the widest range of accretion rates (those for which \citet{GMCPH}
compute a colour-colour diagram position $S_z$)\footnote{We excluded
  4U 1728-34 because this source is thought to accrete from an H-poor 
  donor, and XTE 2123-058 because there are only six bursts in the
  catalogue.}.  We also included two additional sources:  the eclipsing
system EXO 0748-676, which has a large sample of bursts at low
accretion rates that are thought to be triggered by H ignition; and
the accreting millisecond pulsar XTE J1814-338, which has a large
burst sample but where magnetic confinement may also play a role in
setting ignition latitude.   Our simulations can be very easily
extended to neutron stars with different rotation rates:  spreading
speed is inversely proportional to rotation rate, so a more slowly
rotating star will have a higher $v_p$ than a star with more rapid
rotation, all other factors being identical.  The change in oblateness
has a very small effect on convexity and rise time.  

We need to be somewhat cautious in comparing different sources, as
burst properties will differ.  These differences will reflect
variations in the composition
of accreted material, deep crustal heating, and accretion history.
The F diagram for one source may be offset from the F diagram for
another, or different regions of the diagram may be populated,
depending on the source. A star with an H-poor donor, for example,
would not trace out the lower portions of the F diagram.  However, if
our model is correct, then there 
should be two transitional accretion rates for each source, for which ignition
occurs at the pole rather than the equator.  As our simulations show,
not all polar ignition will result in negative convexity bursts.
However, all negative convexity bursts should indicate polar
ignition.  We therefore expect to find negative convexity bursts on
the horizontal strokes of the F diagram.  

\begin{figure*}
\includegraphics[width=18cm, clip]{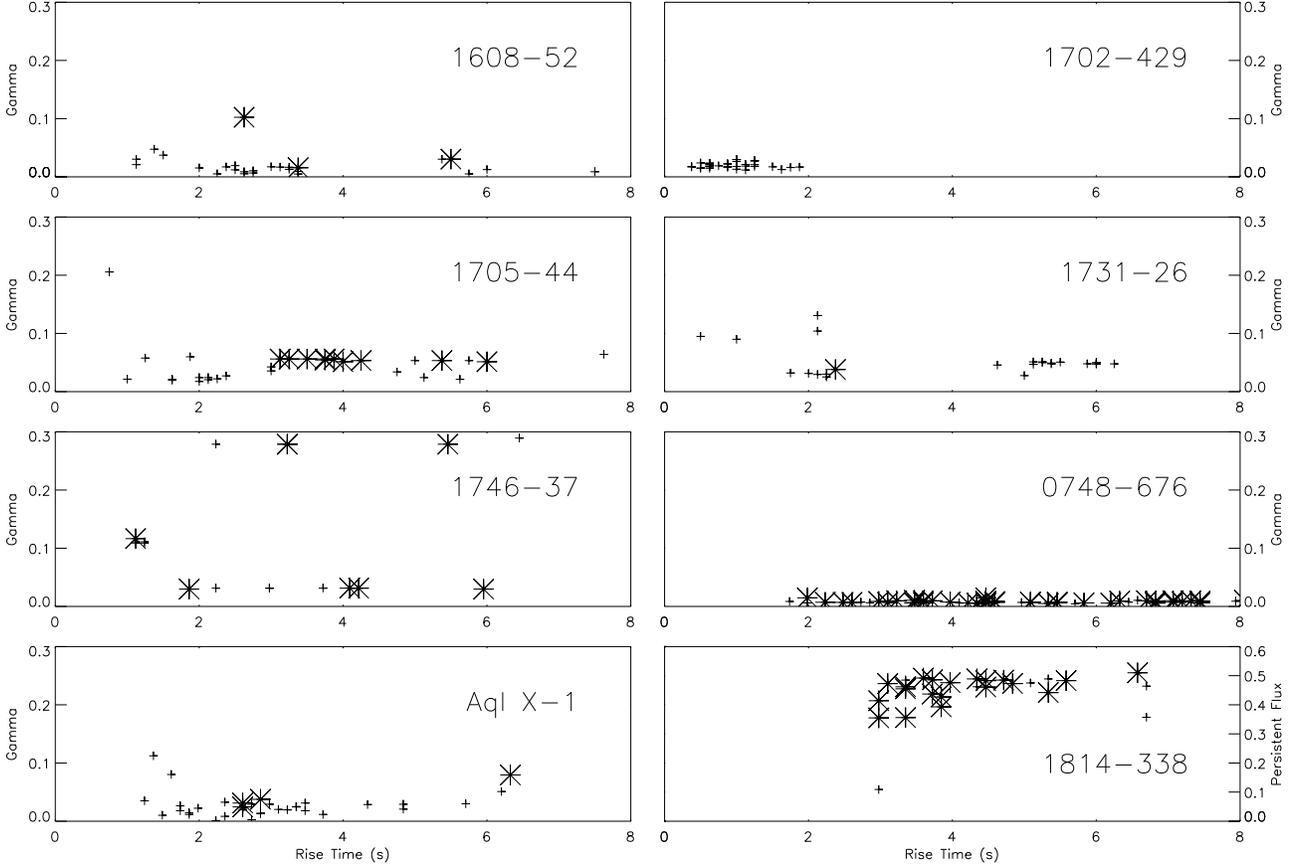}
\caption{Normalized persistent flux ($\gamma$) against burst rise time
  for various burst sources.  Crosses:  bursts with positive convexity.
  Asterisks:  bursts with negative convexity. Rise time is $\tau_R$,
  the time taken for the lightcurve to rise from 10\% to 90\% of the
  peak value.  Values for $\gamma$ (or, in the case of XTE J1814-338,
  persistent flux in units of $10^{-9}$ ergs/cm$^2$/s) are taken from
  \citet{GMCPH}.} 
\label{othersources}
\end{figure*}

In Figure \ref{othersources} we plot F diagrams for the other sources
examined, marking those bursts with negative convexity.  As for 4U
1636-536 we excluded both the faintest bursts and those with kinked or
multi-peaked rises.  For the
accreting millisecond pulsar XTE J1814-338, the
Eddington luminosity has not been measured so we show unscaled
persistent flux rather than $\gamma$.  The properties of the sources
examined are summarized in Table \ref{osrc}.  Inclination is for most
sources unconstrained (although it is expected to be $< 70^\circ$ due
to the absence of dips or eclipses).   The exception is EXO 0748-676,
which is a high inclination eclipsing system.    We expect the range
of accretion rates for which polar ignition is important to be larger
for the more rapidly rotating sources (although whether or not we see
negative convexity bursts would of course depend on the inclination).  

\begin{table*}
\begin{tabular}{l|c|c|c|}
\hline
Source & Frequency (Hz) &
Bursts analyzed (total in catalogue) & Number with $\mathcal{C} < 0$\\
\hline
4U 1608-52 & 619 [1]  &  27 (31) & 3 \\ 
4U 1702-429 & 329 [2] &  38 (44) & 0 \\
4U 1705-44 &   &  30 (39) & 9 \\
KS 1731-26 & 524 [3]  &  24 (27) & 1  \\
4U 1746-37 &  & 20 (30) & 7 \\
EXO 0748-676 & 45 [4]   & 72 (83) & 50  \\
Aql X-1 & 549 [5]  &  35 (40) & 4  \\
XTE J1814-338 & 314 [6]  &  28 (28) & 18  \\
\hline
\end{tabular}
\caption{Properties of the sources depicted in Figure
  \ref{othersources}.  The frequency given is burst oscillation
  frequency except for the pulsar XTE J1814-338, for which the spin
  frequency is known.   References:  [1] \citet{HCG}, [2] \citet{MSS}, [3]
  \citet{SMB}, [4] \citet{VS}, [5] \citet{ZSSS}, [6] \citet{SMSI}  }
\label{osrc}
\end{table*}

The picture that emerges is certainly not clear-cut.  All but one
of the sources (4U 1702-429) show some negative convexity bursts, and many of these sit on what
we might interpret as the lower horizontal stroke of the F diagram.
4U 1746-37 is interesting in that it seems to trace out the entire
upper portion of the F diagram, with two negative convexity bursts at
high accretion rates that might be polar ignition bursts triggered by
He burning.  There are however a few negative convexity bursts at
intermediate accretion rates (as there were for 4U 1636-536).  EXO
0748-676 has a large percentage of negative convexity bursts: if these
are caused by polar ignition then this may seem a little surprising,
since this source is thought to have a much slower spin rate.  This
source does span a very narrow of accretion rates, however, and as a
high inclination system we also expect the south pole ignition bursts
to have negative convexities.  However, if the accretion rate is not
extremely finely-tuned, then there may be some other
factor (such a strong magnetic field) that is controlling fuel
deposition and hence ignition latitude.  The relatively slow spin
inferred for this source suggests that magnetic field effects might be
important.  In this respect the results
from the pulsar (XTE J1814-338) are also extremely
interesting.  This source too shows a large number of negative
convexity bursts, as one might expect if ignition were occurring
off-equator at the magnetic pole.  The phase-locking of persistent
pulsations and burst oscillations in this source also suggests that this
may be the case \citep{SMSI}.  

\section{Summary and conclusions}
\label{conc}

We have shown, using parameterized simulations, that burst rise shape
can be a valuable 
diagnostic of the burning process.  Changes in ignition latitude, in particular, can
have a major impact on burst morphology, and such changes may explain the variation in burst rise shape
seen in the well-studied source 4U 1636-536.  We have argued
that a change from off-equatorial to equatorial ignition
might be the hallmark of the transition from H triggered mixed bursts
to He triggered bursts at low 
accretion rates.   Such a model can also plausibly explain
variations in the detectability of burst oscillations.  There are also areas, however, where additional physics is clearly
required.  Our spreading and burning models were not, for example, able to generate
multi-peaked or kinked burst rises. 

Our goal with this work was to develop a simple phenomenological model
to study the interactions between the various processes operating
during the burst rise, and their influence on lightcurve shape.  What
we have done is clearly simplistic:  the 
various elements of the model will be more closely connected than
we have considered here, and we have made a number of assumptions that
may not be valid.  The parameter space that we have considered,
however, is extremely wide, giving us confidence in our conclusions.
What we have done also demonstrates the power of simple measures of the 
burst shape:  if our model is valid we have been able to identify ignition latitude and
rule out latitude-independent flame spreading speeds, for example (see
also \citet{BSd}).  This type of
study could and should be repeated as more detailed models of the
nuclear burning, spreading and emission process become available.

\section{Acknowledgments}

We would like to thank Edward Brown, Randall Cooper, Andrew Cumming, Fang
Peng, Mike Revnivtsev, Henk Spruit and Rashid Sunyaev and the
anonymous referee for helpful comments.  This research has 
made use of data obtained from 
the High Energy Astrophysics Science Archive Research Centre (HEASARC)
provided by 
NASA's Goddard Space Flight Centre.

\end{document}